\documentclass[twocolumn,showpacs,preprintnumbers,epsfig,prl,aps]{revtex4}
\usepackage{amssymb}

\usepackage{graphicx}
\usepackage{dcolumn}
\usepackage{bm}

\begin{document}

\preprint{}
\title{On the occurrence of Kosterlitz-Thouless behavior in cuprate
superconductors}
\author{T. Schneider}
\affiliation{Physikinstitut, University of Zurich, Winterthurerstrasse 190, 8057 Zurich,
Switzerland. }
\date{\today }

\begin{abstract}
The observation of the characteristic Kosterlitz-Thouless behavior
requires the attaintment of the two dimensional limit\ where the
correlation-length anisotropy, $\gamma =\xi _{ab}/\xi _{c}$,
diverges. Our findings strongly suggest that the failure of several
experiments on films and single crystals to observe any trace of
KT-behavior is attributable either to inhomogeneities or doping by
means of chemical substitution.
\end{abstract}

\pacs{74.20.-z, 64.60.Ak, 74.72.-h}
\maketitle

Since the pioneering work of Kosterlitz and Thouless\cite{kt} (KT)
on the KT transition in the two-dimensional (2D) XY model, much
efforts have been devoted to observe the universal behavior
characteristic of the KT transition, as the universal jump of the
superfluid density\cite{nelson}, measured in $^{4}$He superfluid
films, or the non-linear I - V characteristic, observed in thin
films of conventional superconductors\cite{minnhagen,tsbook}.
Signatures of KT physics can be expected also in layered
superconductors with weak inter-layer coupling. Potential candidates
are underdoped cuprate superconductors where the anisotropy
increases with reduced transition temperature $T_{c}$\cite{tsphysb}.
Recent studies of the I-V characteristic\cite{vadla}, the frequency
dependent conductivity\cite{corson}, the Nernst signal\cite{wang},
the non-linear magnetization\cite{li}, and of the
resistance\cite{matthey} have been interpreted as signatures of KT
behavior. On the other hand, several experiments \cite%
{hosseini,broun,liang,zuev,ruf} failed to observe any trace of the universal
jump in the superfluid density at $T_{c}$.

In this context it is important to recognize that the existence of
the KT-transition (vortex-antivortex dissociation instability) in
$^{4}$He films is intimately connected with the fact that in such
films the interaction energy between vortex pairs depends
logarithmic on the separation between them. As shown by
Pearl\cite{pearl}, vortex pairs in thin superconducting films
(charged superfluid) have a logarithmic interaction energy out to
the characteristic length $\lambda _{2D}=\lambda _{ab}^{2}/d$,
beyond which the interaction energy falls off as $1/r$. Here
$\lambda _{ab}$ is the in-plane penetration depth of the bulk and
$d$ is the film thickness. As $\lambda _{2D}$ increases the
diamagnetism of the superconductor becomes less important.
Consequently, as $\lambda _{2D}$ increases, the vortices in a thin
superconducting film become progressively like those in $^{4}$He
films\cite{beasley} and according to this $\lambda _{2D}>L_{s}=\min
\left[ W,\text{}L\right] $ is required, where $W$ and $L$ denote the
width and the length of the perfect sample. Since real systems, and
in particular the cuprate superconductors are inhomogeneous, the
correlation length $\xi $ cannot grow beyond the lateral extent
$L_{ab}$ of the homogenous regions\cite{bled,tsdan}. This begs the
question of why one should see critical behavior at all if there is
no true phase transition in finite-size systems. The answer is that
critical behavior can be seen in an intermediate temperature regime,
provided that $L_{ab}<L_{s}$ is sufficiently large in order that the
fluctuation dominated regime is attained.

\begin{figure}[tbp]
\centering
\includegraphics[angle=0,width=8.6cm]{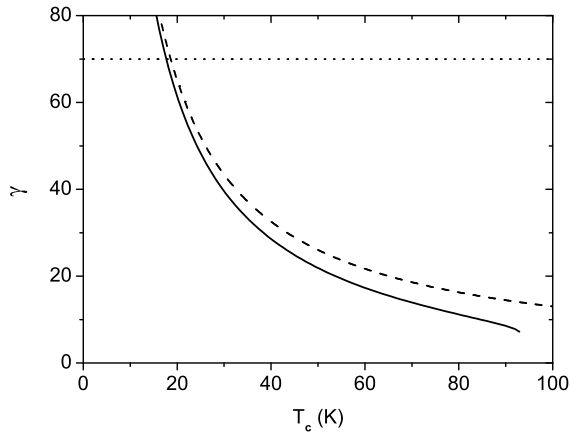}
\caption{$\gamma \left( T_{c}\right) =\gamma \left( T_{cm}\right)
/\left( 1-\left( 1-T_{c}/T_{cm}\right) ^{1/2}\right) $ for
YBa$_{2}$Cu$_{3}$O$_{7-\delta }$ with $\gamma \left( T_{cm}\right)
=7$ and $T_{c}=93$K. The dashed curve is $\gamma \left( T_{c}\right)
=2\gamma \left( T_{cm}\right) T_{cm}/T_{c}$, the leading behavior in
the 2D-limit. The dotted line is $\gamma =70$.} \label{fig1}
\end{figure}

In addition, the occurrence of KT-behavior in single crystals and
thick films of cuprate superconductors requires that the anisotropy
$\gamma =\xi _{ab}/\xi _{c}$ tends to diverge, whereupon the 2D
limit is approached. $\xi _{ab,c}$ denote the correlation length in
the $ab$-plane and along the $c$-axis. In a variety of cuprate
superconductors this behavior is well described by $\gamma \left(
T_{c}\right) =\gamma \left( T_{cm}\right) /\left( 1-\left(
1-T_{c}/T_{cm}\right) ^{1/2}\right) $, where $\gamma \left(
T_{cm}\right) $ denotes the anisotropy at optimum doping where
$T_{c}=T_{cm}$\cite{tsphysb}. In Fig.\ref{fig1} we depicted the
resulting Tc dependence of the anisotropy for
YBa$_{2}$Cu$_{3}$O$_{7-\delta }$. The dashed line is $\gamma \left(
T_{c}\right) =2\gamma \left( T_{cm}\right) /T_{c}$, the limiting
behavior in the 2D-limit. When this limit is attained, a 2D quantum
superconductor to insulator (QSI) transition with dynamic critical
exponent $z=1$ is expected to occur. Here $T_{c}$, the zero
temperature in-plane ($\xi _{ab}\left( 0\right) $) and c-axis ($\xi
_{c}\left( 0\right) $) correlation length, the in-plane magnetic
penetration depth ($\lambda _{ab}\left( 0\right) $) and the
anisotropy scale as $T_{c}\propto 1/\lambda _{ab}^{2}\left( 0\right)
\propto 1/\xi _{ab}\left( 0\right) \propto 1/\gamma $, and
$T_{c}\lambda _{ab}^{2}\left( 0\right) \propto \xi _{c}\left(
0\right) \propto d_{s}$\cite{tsbook,parks,tsphys}. $d_{s}$ denotes
the thickness of the superconducting slabs, becoming independent in
the 2D-limit. In a variety of cuprate superconductors traces of this
behavior, in particular of $T_{c}\propto 1/\lambda _{ab}^{2}\left(
0\right) $, have been observed below $T_{cm}$ in the regime where
the anisotropy scales roughly as $\gamma \left( 0\right) \propto
1/T_{c}$\cite{tsbook,parks,tsphys,uemura}. However, recent
measurements of $\lambda _{c}$\cite{hosseini} and $\lambda
_{ab}$\cite{broun} on YBa$_{2}$Cu$_{3}$O$_{6+x}$ single crystals,
extending to much lower $T_{c}$'s reveal that that the anisotropy
does not attain the 2D-limit when $T_{c}$ is reduced further by
chemical substitution. Their data yields $\gamma \left( T=0\right)
=\lambda _{c}\left( 0\right) /\lambda _{ab}\left( 0\right) \simeq
70$ for $T_{c}$ from $5$ to $15$ K, instead of the characteristic 2D
behavior $\gamma \left( T=0\right) \propto 1/T_{c}$ shown in
Fig.\ref{fig1}. Indeed, their data is consistent with 3D-QSI
critical behavior, namely $T_{c}\propto \lambda _{a,b,c}\left(
0\right) ^{-2z/\left( z+1\right) }\propto \lambda _{a,b,c}\left(
0\right) ^{-1}$\cite{tsbook} with $z\gtrsim 1$\cite{herbut}. $z$
denotes the dynamic critical exponent of the quantum transition.
Consistency with $T_{c}\propto \lambda _{ab}\left( 0\right) ^{-1}$
was also observed by Zuev \textit{et al}.\cite{zuev} in thick
YBa$_{2}$Cu$_{3}$O$_{6+x} $ films.

In contrast, considerable evidence for KT-behavior stems from
resistance and mobile areal carrier density measurements in very
thin (3-4 unit cells thick) underdoped
NdBa$_{2}$Cu$_{3}$O$_{7-\delta }$ films near the 2D-QSI transition
using the electric field effect technique\cite{matthey}. To estimate
the $T_{c}$'s, ranging from $0.5$ to $10$K, consistency with the
characteristic KT-behavior of the resistance, $\rho =\rho _{0}\exp
(-bt^{-1/2})$\cite{kt}, was established, where $t=T/T_{c}-1$ and
$\rho _{0}$ and $b$ are material dependent parameters. Furthermore
the measurements of the electric field induced changes of $T_{c}$
and the areal carrier density n$_{2D}$ uncovered the relationship,
$T_{c}\propto n_{2D}^{z\overline{\nu }}$ with $z\overline{\nu }=1$,
the signature of a 2D-QSI-transition\cite{herbut2,herbut3}.
$\overline{\nu }$ is the critical exponent of the zero temperature
in-plane penetration depth, $\xi _{ab}\left( 0\right) \propto
n_{2D}^{\overline{\nu }}$. Together with the quantum counterpart of
the Nelson -Kosterlitz (NK)-relation\cite{nelson} it implies
$T_{c}\propto n_{2D}\propto 1/1/\lambda _{ab}^{2}\left( 0\right) $,
characteristic for a 2D-QSI transition with $z\overline{\nu }=1$. In
this context it is interesting to note that a magnetic field tuned
2D-QSI transition with $z\overline{\nu }=1.37\pm 0.1$ was also
observed in YBa$_{2}$Cu$_{3}$O$_{7-\delta }$ single crystals with
$T_{c}\approx 2$K\cite{seidler}. Apparently, the reduction of
$T_{c}$ by means of chemical substitution does not fully decouple
the superconducting sheets. This provides a key anchor point for the
understanding of the phase diagram of underdoped cuprate
superconductors - distinct quantum critical points in the chemical
substitution and electric field effect or magnetic field tuned case.

\begin{figure}[tbp]
\centering
\includegraphics[angle=0,width=8.6cm]{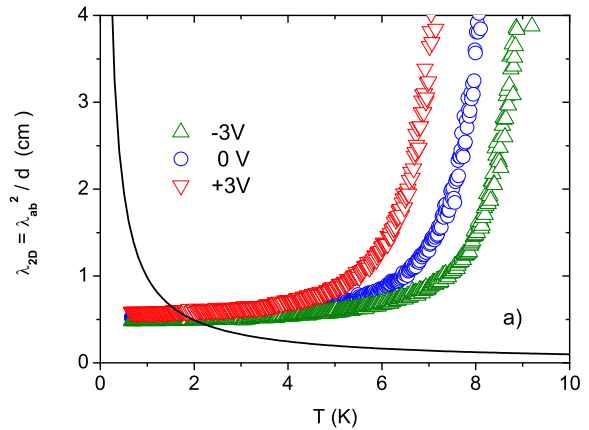}
\includegraphics[angle=0,width=8.6cm]{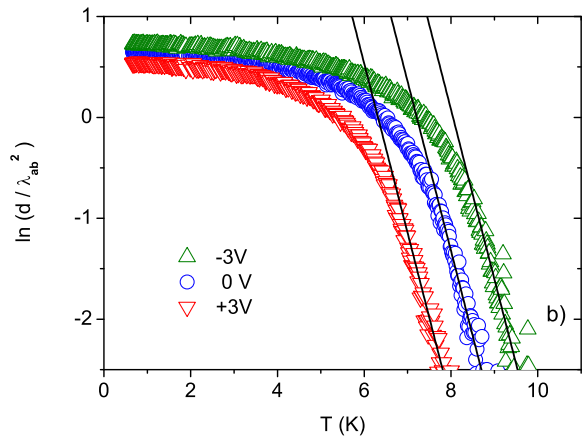}
\caption{a) $\protect\lambda _{2D}=\protect\lambda _{ab}^{2}/d$
\textit{vs}. $T$ derived from the data of R\"{u}fenacht
\textit{et.al}. \protect\cite{ruf} \ The solid line is the
Nelson-Kosterlitz-line given by Eq.(\protect\ref{eq1}). b) ln$\left(
d/\protect\lambda _{ab}^{2}\right) $ \textit{vs}. $T$ derived from
the data of R\"{u}fenacht \textit{et.al}. \protect\cite{ruf} The
solid lines are Eq.(\protect\ref{eq2}) with ln$\left( A\right)
=13.5$, $2.1$, $10.6$ and $BL_{ab}=1.68$ (K$^{-1}$) for the gate
voltages $V=-3$, $0$, and $+3$ (V), respectively.} \label{fig2}
\end{figure}

Finally we turn to the measurements of R\"{u}fenacht \textit{et
al}.\cite{ruf}. They reported the temperature and electric-field
dependence of $d/\lambda _{ab}^{2}$, obtained by capacitively
charging an epitaxially grown ultrathin (two-unit-cell-thick) LSCO
film in the underdoped regime ($x\simeq 0.1$) with an electrostatic
field applied across a gate insulator with a high dielectric
constant. In Fig.\ref{fig2}a we show $\lambda _{2D}=\lambda
_{ab}^{2}/d$ \textit{vs}. $T$ derived from their data. The solid
line is the Nelson-Kosterlitz line,
\begin{equation}
k_{B}T_{c}=\frac{\pi }{2}\frac{\Phi _{0}^{2}}{16\pi
^{3}}\frac{d}{\lambda _{ab}^{2}\left( T_{c}^{-}\right) },
\label{eq1}
\end{equation}
where $d/\lambda _{ab}^{2}\left( T\right) $ jumps to zero at
$T_{c}^{-}$ \cite{nelson}. In view of the lateral extent of the
film, $0.25$ cm, and the expected KT-transition around $T\simeq
1.8$K, the condition, $\lambda _{2D}>> $ $\min \left[ W,\text{
}L\right] $, for the occurrence of KT-behavior is fairly satisfied.
In a thin-film superconductor vortices interact logarithmically out
to a distance on the order of $\lambda _{2D}$, at which point the
interaction approaches a constant. Thus, because $\lambda
_{2D}\left( T\simeq 1.8K\right) $ exceeds the lateral extent of the
film, the system does not exhibit what amounts to an intrinsic
finite-size effect\cite{pierson}. Nevertheless, there is no
signature of the characteristic jump. Instead there is an extended
tail which appears to preclude a sharp transition at higher
temperatures. From the studies of finite 2D-XY-models it is known
that the superfluid density remains finite above the transition
temperature of the infinite counterpart. More precisely, it
decreases smoothly and the tail increases with reduced lateral
system size $L_{ab}$ \cite{schlutka,harada,minnhagenkim}. If the
pronounced tails observed by R\"{u}fenacht \textit{et al}.\cite{ruf}
are attributable to an inhomogeneity induced finite size effect due
to the limited extent $L_{ab}$ of the homogeneous regions, their
temperature dependence should be of the form\cite{weber}
\begin{equation}
\frac{d}{\lambda _{ab}^{2}\left( T\right) }\simeq A\exp \left(
-BTL_{ab}\right) .  \label{eq2}
\end{equation}%
A glance to Fig.\ref{fig2}b, showing ln$\left( d/\lambda
_{ab}^{2}\right) $ \textit{vs}. $T$, reveals that the observed tails
are remarkably consistent with this finite size behavior and a
unique value for $BL_{ab}$. Apparently, the field induced modulation
of $d/\lambda _{ab}^{2}$ and $\lambda _{2D}=\lambda _{ab}^{2}/d$
does not affect $L_{ab}$. This confirms that the rounding stems from
an inhomogeneity induced finite size effect.

To summarize, our findings strongly suggest that the failure of
several experiments\cite{hosseini,broun,liang,zuev,ruf} to observe
any trace of the universal jump in the temperature dependence of the
superfluid density is attributable to the failure to attain the 2D
limit. While in YBa$_{2}$Cu$_{3} $O$_{6+x}$ single crystals and
thick films chemical substitution makes it impossible to attain this
limit, the observation of KT-behavior in ultrathin films requires
that the 2D penetration depth $\lambda _{2D}=\lambda _{ab}^{2}/d$
exceeds near the Nelson-Kosterlitz line the lateral extent of the
films and unique homogeneity. Furthermore, the observation of
distinct quantum superconductor to insulator transitions in in the
chemical substitution and electric field effect or magnetic field
tuned case, provides a key anchor point for the understanding of the
phase diagram of underdoped cuprate superconductors.

I would like to thank J.-M. Triscone and N. Reyren for useful discussions
and A. R\"{u}fenacht for providing the data used in Fig. \ref{fig2}.

\end{document}